\documentclass{article}

 \usepackage[main, final]{neurips_2025}

\usepackage[utf8]{inputenc} 
\usepackage[T1]{fontenc}    
\usepackage{hyperref}       
\usepackage{url}            
\usepackage{booktabs}       
\usepackage{amsfonts}       
\usepackage{nicefrac}       
\usepackage{microtype}      
\usepackage{xcolor}         

\usepackage{amsmath}
\usepackage[capitalize,noabbrev]{cleveref}
\usepackage{amssymb}
\usepackage{mathtools}
\usepackage{amsthm}
\usepackage{xcolor}
\usepackage{sverb, longtable}
\usepackage{rotating}
\usepackage{epstopdf}
\DeclareMathOperator*{\argmax}{arg\,max}

\usepackage{multirow}%
\usepackage{mathrsfs}%
\usepackage{booktabs}%
\usepackage{bm}
\usepackage{tikz}
\usepackage{subcaption}
\usepackage{wrapfig}

\usepackage{algorithm}
\usepackage{algpseudocode}

\usetikzlibrary{calc}
\usetikzlibrary{arrows,shapes,chains}
\usetikzlibrary{decorations.pathmorphing}
\usepgflibrary{patterns}

\title{DoDo-Code: an Efficient Levenshtein Distance Embedding-based Code for 4-ary IDS Channel}

\author{%
    Alan J.X. Guo\textsuperscript{\rm 1,\,2}\thanks{Corresponding author.}, ~Sihan Sun\textsuperscript{\rm 1}, ~Xiang Wei\textsuperscript{\rm 1}, ~Mengyi Wei\textsuperscript{\rm 1}, ~Xin Chen\textsuperscript{\rm 1,\,2}\\
    \textsuperscript{\rm 1}Center for Applied Mathematics, KL-AAGDM, 
    Tianjin University, 
    Tianjin 300072, China\\
    \textsuperscript{\rm 2}State Key Laboratory of Synthetic Biology, 
    Tianjin University, 
    Tianjin 300072, China\\
  \texttt{\{jiaxiang.guo, sihansun, weixiang, mengyi.wei, chen\_xin\}@tju.edu.cn} \\
}

\begin{document}

\maketitle

\begin{abstract}
    With the emergence of new storage and communication methods, 
    the insertion, deletion, and substitution (IDS) channel has attracted considerable attention.
    However, many topics on the IDS channel and the associated Levenshtein distance remain open, 
    making the invention of a novel IDS-correcting code a hard task. 
    Furthermore, current studies on single-IDS-correcting code misalign with the requirements of applications 
    which necessitates the correcting of multiple errors. 
    Compromise solutions have involved shortening codewords to reduce the chance of multiple errors. 
    However, the code rates of existing codes are poor at short lengths, diminishing the overall storage density. 
    In this study, a novel method is introduced for designing high-code-rate single-IDS-correcting codewords 
    through deep Levenshtein distance embedding. 
    A deep learning model is utilized to project the sequences into embedding vectors 
    that preserve the Levenshtein distances between the original sequences. 
    This embedding space serves as a proxy for the complex Levenshtein domain, 
    within which algorithms for codeword search and segment correcting is developed. 
    While the concept underpinning this approach is straightforward, 
    it bypasses the mathematical challenges typically encountered in code design. 
    The proposed method results in a code rate that outperforms existing combinatorial solutions, 
    particularly for designing short-length codewords.
\end{abstract}

\section{Introduction}


With the emergence of new storage and communication methods~\cite{church2012next,goldman2013towards,grass2015robust,organick2018random,shaikh2022high}, 
insertion, deletion, and substitution (IDS) channels over non-binary symbols have attracted significant attention. 
However, applying existing IDS-correcting codes to practical applications is not straightforward and 
faces challenges such as a low overall code rate and limited multiple error-correcting capabilities.

Most of the current IDS-correcting codes focus on correcting 
a single error~\cite{varshamov1965code,levenshtein1966binary,sloane2000single,cai2021correcting,garbys2023beyond} 
or a burst of errors~\cite{schoeny2017codes,khuat2021quaternary,wang2023non}, 
with an emphasis on achieving asymptotic optimality in code rate. 
These codes are typically varieties of the Varshamov-Tenengolts (VT) 
code~\cite{varshamov1965code,calabi1969family,tanaka1976synchronization,tatwawadi2019tutorial}. 
Codebook generation has recently gained attention in specific tasks, such as 
the $p$-substitution-$k$-deletion code~\cite{lu2024unrestricted}.

However, in most applications, the ability to correct multiple IDS errors is crucial. 
This is because IDS errors may occur simultaneously at different positions along the codeword, and 
the likelihood of insertion, deletion, and substitution can vary due to channel properties, 
making them unequal in occurrence~\cite{blawat2016forward,welter2024end}. 

To address the challenge of correcting multiple errors, 
a compromise approach is to employ segmented error-correcting code, 
as proposed in previous studies~\cite{liu2009codes,abroshan2017coding,cai2021coding,yan2022segmented,maarouf2022concatenated,welter2024end}. 
The sequence is implicitly segmented into disjoint segments, each capable of correcting one single 
IDS error. 
This segmentation enables the sequence to rectify multiple errors to some extent. 
However, the possibility remains that multiple errors may occur within the same segment. 
Researchers can employ an outer error-correcting code (ECC) to address such failures. 

Unfortunately, the segmented error-correcting codes usually have low code rate, since 
these codes use the same code rate as their underlying single-IDS-correcting code within each segment, 
and current IDS-correcting codes have low code rate for small codeword lengths. 
For example, an order-optimal code of length $n$ over the $4$-ary alphabet using 
$\log{n} + O(\log\log{n})$ redundancy bits is introduced in~\cite{cai2021correcting}. 
The state-of-the-art code rate till now is proposed in~\cite{garbys2023beyond} 
with redundancy bits of $\log n+\log\log n+7+o(1)$, 
which reduces the redundancy by $6$ bits compared to the original code in~\cite{cai2021correcting}. 
Although such codes are efficient when $n$ is large, the constant term in the number of redundancy bits 
limits their code rate when $n$ is small. 
Given this, there is potential to enhance the code rate of segmented error-correcting codes, because the segments usually 
use shorter codewords~\cite{yan2022segmented}. 

This work focus on constructing a $4$-ary code that uses fewer redundancy bits 
than the existing order-optimal code offered by the mathematicians~\cite{cai2021correcting,garbys2023beyond} 
at the end of short-length codewords. 
Namely, 
following the bounded distance decoder (BDD) which is one of the basics of classical codes~\cite{richardson2008modern}, 
the novel DoDo-Code is proposed
by leveraging the deep embedding of the Levenshtein distance. 
The embedding space is utilized as a proxy for the intricate Levenshtein domain, facilitating high code rate codebook design 
and fast IDS-correcting. 
The proposed DoDo-Code uses a comprehensive approach that includes the following key procedures: 
deep embedding of the Levenshtein domain, 
deep embedding-based greedy search of codewords, 
and deep embedding-based segment correcting. 


As a result, the proposed DoDo-Code offers a solution for IDS-correcting code when code length is short, 
excelling in the following aspects: 
\begin{itemize}
\item The proposed DoDo-Code achieves a code rate that surpasses the state-of-the-art 
and shows characteristics of ``optimality'' when code length $n$ is small. 
\item With one edit operation corrupted codewords can be firmly corrected, 
the computational complexity is effectively reduced to $O(n)$ of the decoder to correct IDS errors. 
\end{itemize}
To the best of our knowledge, the proposed DoDo-Code is the first IDS-correcting code designed by deep learning methods 
and the first IDS-correcting code that shows characteristics of ``optimality'' when the code length is small. 

\section{Related works}
While neural channel codes~\cite{jamali2022productae,choukroun2022error,chahine2023deepic,saber2024generalized,choukroun2024learning,park2025crossmpt} 
have recently gained attention and achieved state-of-the-art performance in several settings, 
they offer little help in addressing the aforementioned issue. 
This is because they primarily target the additive white Gaussian noise (AWGN) channel 
and handle flip or erasure errors, rather than IDS errors. 
Moreover, being neural network-based, 
these solutions typically fail to generate the static, explicit codebooks 
for downstream applications.

\section{Bounded Distance Decoder}
The proposed code employs a quite fundamental approach called the BDD; 
let's revisit the basics of classical codes~\cite{richardson2008modern}. 

Given a code $C$, which is a collection of codewords, its minimum distance is defined as follows: 
\begin{equation}
    d(C) = \min\{d(\bm{c}_i,\bm{c}_j):\bm{c}_i,\bm{c}_j\in C,i\neq j\}.
\end{equation}
Once a code $C$ with a minimum distance of $d$ is constructed, 
a BDD can be deployed with a decoding radius $r=\lfloor\frac{d-1}{2}\rfloor$ by
correcting a corrupted word $\hat{\bm{c}}$ to the corresponding codeword $\bm{c}$ such that $d(\hat{\bm{c}},\bm{c})\leq r$. 

In the context of correcting IDS errors, the Levenshtein distance~\cite{levenshtein1966binary} plays a pivotal role. 
It is defined as the minimum number of insertions, deletions, and substitutions required to transform 
one sequence into another. 
According to the principles of BDD, to correct a single IDS error, the decoding radius in terms of Levenshtein distance 
is $r=\lfloor\frac{d-1}{2}\rfloor=1$, 
meaning the minimum distance of the code must be at least $d=3$. 
Therefore, constructing a code $C(n)$ for a fixed code length $n$, 
whose elements have mutual Levenshtein distances greater than $3$, is considered. 

However, the construction of such a code faces three significant challenges.
Firstly, the sizes of Levenshtein balls, representing the set of a sequence and 
its neighboring sequences within a Levenshtein distance of $r$, 
exhibit a lack of homogeneity~\cite{barlev2023size,wang2022size}. 
Researchers want to select sequences with small Levenshtein balls as the codewords to enhance the code rate, 
but a clear depiction of Levenshtein balls is still absent~\cite{barlev2023size,wang2022size}. 
An algorithm surpassing random codeword selecting has not yet been published to the best of our knowledge. 
Secondly, the computational complexity of the Levenshtein distance-based BDD is substantial. 
The complexity of computing the Levenshtein distance is at least 
$O(n^{2-\epsilon}),\forall \epsilon>0$~\cite{masek1980faster,backurs2015edit}, 
unless the strong exponential time hypothesis is false. 
Additionally, most existing neighbor searching algorithms, which are keys to the BDD process, 
are primarily designed for conventional distance~\cite{cunningham2021nearest} 
and inapplicable to the Levenshtein distance. 

\section{DoDo-Code}
\subsection{Deep embedding of Levenshtein distance}\label{subsec:dsee}
Considering the complexity of calculating the Levenshtein distance, 
researchers have explored mapping sequences into embedding vectors,
using a conventional distance between these vectors to approximate the Levenshtein distance~\cite{ostrovsky2007low,chakraborty2016streaming}.  
Recently, deep learning techniques have been employed for Levenshtein distance embedding 
and achieved remarkable performance across various works~\cite{koide2018neural,zhang2019neural,dai2020convolutional,corso2021neural,guo2022deep,wei2023levenshtein}.

From a broader perspective, it is found that these embeddings not only accelerate the Levenshtein distance estimation, 
but also offer a way to analyze the properties and structures of the Levenshtein distance. 
The embedding aims to create a vector space where the squared Euclidean distance between vectors serves 
as a proxy for the Levenshtein distance between the original sequences. 
This allows us to leverage the geometric structure of the embedding space to reason about the complex 
combinatorial properties of the Levenshtein domain. 
For example, in the vector space, the embedding vectors of sequences within a Levenshtein ball 
should naturally exhibit a tight clustering. 
The embedding model from~\cite{wei2023levenshtein} is modified 
to focus more on the Levenshtein neighbor relations between the sequences in our work. 

Let $\bm{s}$ and $\bm{t}$ denote two sequences of length $n$ on the alphabet $\{0,1,2,3\}$, 
and let $d = \Delta_L(\bm{s},\bm{t})$ represent the groundtruth Levenshtein distance between them. 
Our task is to identify an embedding function $f(\cdot)$, such that the mapped embedding vectors $\bm{u}=f(\bm{s})$ 
and $\bm{v}=f(\bm{t})$ have a conventional distance $\hat{d}= \Delta(\bm{u},\bm{v})$ approximates to the groundtruth 
Levenshtein distance $d=\Delta_L(\bm{s},\bm{t})$. 
Let the embedding function $f(\cdot;\theta)$ be a deep embedding model with learnable parameters $\theta$, 
which is implemented as a model with $10$ $1$D-CNN~\cite{krizhevsky2012imagenet} layers 
and one final batch normalization~\cite{ioffe2015batch} in this study. 
The training of $f(\cdot;\theta)$ can be expressed as an optimization of:
\begin{eqnarray}\label{eqn:loss}
	\hat{\theta} &=& \mathop{\arg\min}_{\theta} \sum \mathcal{L}(d,\hat{d};\theta) \\
	&=& \mathop{\arg\min}_{\theta} \sum \mathcal{L}(d,\Delta(f(\bm{s};\theta),f(\bm{t};\theta))),
\end{eqnarray}
where the function $\mathcal{L}(\cdot,\cdot)$ is a predefined loss function that 
measures the disparity between the predicted distance and the groundtruth distance. 
By optimizing \eqref{eqn:loss}, the parameters of the embedding model are learned and denoted as $\hat{\theta}$, 
and the optimized deep embedding model $f(\cdot;\hat{\theta})$ is capable of mapping sequences 
to their corresponding embedding vectors. 
This model configuration is often referred to as a Siamese neural network~\cite{bromley1993signature}. 
A brief illustration of the utilized Siamese neural network is presented in \cref{fig:siamese}. 

The squared Euclidean distance between the embedding vectors 
is employed as the approximation of the groundtruth Levenshtein distance. 
It is defined as: 
\begin{equation}
	\hat{d}= \Delta(\bm{u},\bm{v}) = \sum_{i}(u_i-v_i)^2.
\end{equation}
Although the squared Euclidean distance is not a true distance metric, its effectiveness has been validated in~\cite{guo2022deep}. 
Moreover, since it is simply the square of the Euclidean distance, 
it remains compatible with most neighbor searching algorithms in Euclidean space.  

\begin{wrapfigure}{r}{0.5\textwidth}
    \vskip -0.1in
    \centering
{\linespread{1}
		\centering
		\tikzstyle{format}=[circle,draw,thin,fill=white]
		\tikzstyle{format_gray}=[circle,draw,thin,fill=gray]
		\tikzstyle{format_rect}=[rectangle,draw,thin,fill=white,align=center]
		\tikzstyle{arrowstyle} = [->,thick]
		\tikzstyle{network} = [rectangle, minimum width = 3cm, minimum height = 1cm, text centered, draw = black,align=center,rounded corners,fill=green_so,fill opacity=0.5,text opacity=1]
		\tikzstyle{training_batch} = [trapezium, trapezium left angle = 30, trapezium right angle = 150, minimum width = 3cm, text centered, draw = black, fill = cyan_so, fill opacity=0.3,text opacity=1,align=center]		
		\tikzstyle{class_features} = [trapezium, trapezium left angle = 30, trapezium right angle = 150, minimum width = 3cm, text centered, draw = black, fill = cyan_so, fill opacity=0.3,text opacity=1,align=center]
		\tikzstyle{pixel} = [rectangle, draw = black, fill = orange_so, fill opacity=0.5,text opacity=0,align=center]	
		\tikzstyle{pixel_red} = [rectangle, draw = black, fill = red_so, fill opacity=1,text opacity=0,align=center]	
		\tikzstyle{feature} = [rectangle, draw = black, fill = orange_so, fill opacity=0.3,text opacity=0,align=center,rounded corners]	
		\tikzstyle{feature_sfp} = [rectangle, draw = black, fill = violet_so, fill opacity=0.3,text opacity=0,align=center,rounded corners]					
		\tikzstyle{arrow1} = [thick, ->, >= stealth]
		\tikzstyle{arrow1_thick} = [thick, ->, >= stealth, line width=1.2pt]
		\tikzstyle{arrow2} = [thick, dashed, ->, >= stealth]
		\tikzstyle{thick_line} = [thick, line width=1.5pt]
		\tikzstyle{channel} = [fill=white,fill opacity = 0.7, rounded corners=3pt]
		\tikzstyle{channel_shadow} = [fill = gray_so, fill opacity = 0.1, rounded corners]
		\tikzstyle{channel_selected} = [fill = orange_so, fill opacity = 0.5]

		\tikzstyle{dna} = [decoration={coil}, decorate, thick, decoration={aspect=0, segment length=0.5*0.87cm, post length=0.,pre length=0.}]
		\scalebox{0.77} 
		{
		\begin{tikzpicture}[auto,>=latex', thin, start chain=going below, every join/.style={norm}]
				\definecolor{gray_so}{RGB}{88,110,117}
				\definecolor{lightgray_so}{RGB}{207,221,221}
				\definecolor{yellow_so}{RGB}{181,137,0}
				\definecolor{cyan_so}{RGB}{42,161,152}
				\definecolor{orange_so}{RGB}{203,75,22}
				\definecolor{green_so}{RGB}{133,153,0}
				\definecolor{red_so}{RGB}{220,50,47}
				\definecolor{magenta_so}{RGB}{211,54,130}
				\definecolor{violet_so}{RGB}{108,113,196}
				\definecolor{yellow_ad}{RGB}{242,228,201}
				\definecolor{pink_ad}{RGB}{242,182,160}
				\definecolor{green_ad}{RGB}{146,195,185}
				\definecolor{dgreen_ad}{RGB}{104,166,148}
				\definecolor{purple_ad}{RGB}{115,72,88}
				\definecolor{att_blue}{RGB}{185,233,248}
				\definecolor{nature_orange}{RGB}{252,140,98}
				\definecolor{nature_green}{RGB}{102,195,170}
				\definecolor{nature_blue}{RGB}{142,160,204}
				\definecolor{nature_yellow}{RGB}{253,184,55}
				\useasboundingbox  (0,0) rectangle (8,5.2);

				\scope[transform canvas={scale=1}]
				
				\coordinate (zero) at (0,0);
				\coordinate (upperhalf) at (0,4);
				\coordinate (lowerhalf) at (0,1.2);

				\coordinate (upperDnaCenter) at ($(upperhalf)+(0.8,0)$);
				\coordinate (lowerDnaCenter) at ($(lowerhalf)+(0.8,0)$);

				\draw[dna, decoration={amplitude=.15cm}] ($(upperDnaCenter)+(0,-0.95)$) -- ($(upperDnaCenter)+(0,1.02)$);
				\draw[dna, decoration={amplitude=-.15cm}] ($(upperDnaCenter)+(0,-1.02)$) -- ($(upperDnaCenter)+(0,0.95)$);
				\node at ($(upperDnaCenter)+(-0.4,0.7)$) {$\bm{s}$};

				\draw[dna, decoration={amplitude=.15cm}] ($(lowerDnaCenter)+(0,-0.95)$) -- ($(lowerDnaCenter)+(0,1.02)$);
				\draw[dna, decoration={amplitude=-.15cm}] ($(lowerDnaCenter)+(0,-1.02)$) -- ($(lowerDnaCenter)+(0,0.95)$);
				\node at ($(lowerDnaCenter)+(-0.4,0.7)$) {$\bm{t}$};

				\coordinate (upperModelCenter) at ($(upperDnaCenter)+(2.2,0)$);
				\coordinate (lowerModelCenter) at ($(lowerDnaCenter)+(2.2,0)$);
				\filldraw[channel, fill=nature_orange] ($(upperModelCenter)+(-1,-0.618)$) rectangle ($(upperModelCenter)+(1,0.618)$);
				\node at ($(upperModelCenter)+(0,0.33)$) {\small{$f(\cdot;\theta):$}};
				\node at ($(upperModelCenter)+(0,-0.0)$) {\small{Embedding}};
				\node at ($(upperModelCenter)+(0,-0.33)$) {\small{Network}};
				\filldraw[channel, fill=nature_orange] ($(lowerModelCenter)+(-1,-0.618)$) rectangle ($(lowerModelCenter)+(1,0.618)$);
				\node at ($(lowerModelCenter)+(0,0.33)$) {\small{$f(\cdot;\theta):$}};
				\node at ($(lowerModelCenter)+(0,-0.0)$) {\small{Embedding}};
				\node at ($(lowerModelCenter)+(0,-0.33)$) {\small{Network}};
				\draw[arrow1_thick] ($(upperDnaCenter)+(0.25,0)$) -- ($(upperModelCenter)+(-1.,0)$);
				\draw[arrow1_thick] ($(lowerDnaCenter)+(0.25,0)$) -- ($(lowerModelCenter)+(-1.,0)$);

				\coordinate (weightsCenter) at ($(upperModelCenter)!0.5!(lowerModelCenter)$);
				\filldraw[channel, fill=white, dashed, double] ($(weightsCenter)+(-1,-0.382)$) rectangle ($(weightsCenter)+(1,0.382)$);
				\node at ($(weightsCenter)+(0,0.17)$) {\small{Shared}};
				\node at ($(weightsCenter)+(0,-0.17)$) {\small{Weight}};
				\draw[arrow1, rounded corners, double,line width=0.7pt, dashed] ($(weightsCenter)+(0,0.382+0.)$) -- ($(upperModelCenter)+(0,-0.618-0.)$);
				\draw[arrow1, rounded corners, double,line width=0.7pt, dashed] ($(weightsCenter)+(0,-0.382-0.)$) -- ($(lowerModelCenter)+(0,0.618+0.)$);

				\coordinate (upperFeatureCenter) at ($(upperModelCenter)+(2.2,0)$);
				\coordinate (lowerFeatureCenter) at ($(lowerModelCenter)+(2.2,0)$);
				\filldraw[channel_selected,fill=white] ($(lowerFeatureCenter)+(-0.125,-0.9)$) rectangle ($(lowerFeatureCenter)+(0.125,0.9)$);
				\filldraw[channel_selected,fill=nature_yellow,fill opacity = 0.5] ($(lowerFeatureCenter)+(0,0)+(-0.125,-0.9)$) rectangle ($(lowerFeatureCenter)+(0.25,0.25)+(-0.125,-0.9)$);
				\filldraw[channel_selected,fill=nature_blue,fill opacity = 0.5] ($(lowerFeatureCenter)+(0,1.55)+(-0.125,-0.9)$) rectangle ($(lowerFeatureCenter)+(0.25,1.8)+(-0.125,-0.9)$);
				\filldraw[channel_selected,fill=nature_green,fill opacity = 1] ($(lowerFeatureCenter)+(0,1.3)+(-0.125,-0.9)$) rectangle ($(lowerFeatureCenter)+(0.25,1.55)+(-0.125,-0.9)$);
				\filldraw[channel_selected,fill=nature_orange,fill opacity = 1] ($(lowerFeatureCenter)+(0,1.05)+(-0.125,-0.9)$) rectangle ($(lowerFeatureCenter)+(0.25,1.30)+(-0.125,-0.9)$);

				\filldraw[channel_selected,fill=white] ($(upperFeatureCenter)+(-0.125,-0.9)$) rectangle ($(upperFeatureCenter)+(0.125,0.9)$);
				\filldraw[channel_selected,fill=nature_yellow,fill opacity = 1] ($(upperFeatureCenter)+(0,0)+(-0.125,-0.9)$) rectangle ($(upperFeatureCenter)+(0.25,0.25)+(-0.125,-0.9)$);
				\filldraw[channel_selected,fill=nature_blue,fill opacity = 1] ($(upperFeatureCenter)+(0,1.55)+(-0.125,-0.9)$) rectangle ($(upperFeatureCenter)+(0.25,1.8)+(-0.125,-0.9)$);
				\filldraw[channel_selected,fill=nature_green,fill opacity = 0.5] ($(upperFeatureCenter)+(0,1.3)+(-0.125,-0.9)$) rectangle ($(upperFeatureCenter)+(0.25,1.55)+(-0.125,-0.9)$);
				\filldraw[channel_selected,fill=nature_orange,fill opacity = 0.5] ($(upperFeatureCenter)+(0,1.05)+(-0.125,-0.9)$) rectangle ($(upperFeatureCenter)+(0.25,1.30)+(-0.125,-0.9)$);
				\node at ($(upperFeatureCenter)+(-0.4,0.7)$) {$\bm{u}$};
				\node at ($(lowerFeatureCenter)+(-0.4,0.7)$) {$\bm{v}$};

				\draw[arrow1_thick] ($(upperModelCenter)+(1.,0)$) -- ($(upperFeatureCenter)+(-0.125-0.,0)$);
				\draw[arrow1_thick] ($(lowerModelCenter)+(1.,0)$) -- ($(lowerFeatureCenter)+(-0.125-0.,0)$);

				\coordinate (upperEuclideanCenter) at ($(upperFeatureCenter)+(1.5,0)$);
				\coordinate (lowerEuclideanCenter) at ($(lowerFeatureCenter)+(1.5,0)$);
				\coordinate (euclideanCenter) at ($(upperEuclideanCenter)!0.5!(lowerEuclideanCenter)$);

				\filldraw[channel, dashed,line width=1.2pt] ($(euclideanCenter)+(-1,-0.382)$) rectangle ($(euclideanCenter)+(1,0.382)$);
				\node at ($(euclideanCenter)+(0,0)$) {$\hat{d} = \Delta(\bm{u},\bm{v})$};

				\draw[arrow1_thick, rounded corners, dashed] ($(upperFeatureCenter)+(0.125+0.,0)$) -- ($(upperEuclideanCenter)+(0,0)$) -- ($(euclideanCenter)+(0,0.4+0.)$);
				\draw[arrow1_thick, rounded corners, dashed] ($(lowerFeatureCenter)+(0.125+0.,0)$) -- ($(lowerEuclideanCenter)+(0,0)$) -- ($(euclideanCenter)+(0,-0.4-0.)$);

				\coordinate (outputCenter) at ($(euclideanCenter)+(2.2,0)$);
				\endscope
		\end{tikzpicture}	
		}
}
\caption{\label{fig:siamese} Siamese neural network. 
Given two sequences $\bm{s},\bm{t}$, 
mapped to respective embedding vectors
$\bm{u}, \bm{v}$. 
The approximated distance is calculated as a conventional distance between 
$\bm{u}$ and $\bm{v}$. 
}
\vskip -0.in
\end{wrapfigure}

The negative log-likelihood loss with the Poisson distribution (PNLL)~\cite{wei2023levenshtein}, which is formulated as: 
\begin{equation}\label{eqn:pnll}
    \mathcal{L}(d,\hat{d})=\mathrm{PNLL}(\hat{d};d) = \hat{d} - d\ln{\hat{d}},
\end{equation}
has been proposed to provide a global approximation 
of the Levenshtein distance. 
In this work, 
the code's construction is dependent on local structures within the Levenshtein distance domain, 
namely sequences at a distance of $1$ or $2$. 
Consequently, from the perspective of the greedy search algorithm, there is no functional difference 
between using the complete Levenshtein distance and a truncated version. 
However, this distinction is raised for training the embedding model, 
relaxing the optimization objective to a truncated Levenshtein distance is an easier learning task 
for the model than approximating the global Levenshtein distance metric precisely. 
In view of this, the loss function is revised to emphasize the approximations between sequence pairs 
within the Levenshtein balls of radius $2$. 
Specifically, the model is trained to provide a precise prediction for Levenshtein distance $1$ and 
to ensure that the predicted distance is greater than $2$ when the groundtruth distance is greater or equal to $2$. 
The revised loss function is defined as follows:
\begin{equation}\label{eqn:revisedpnll}
	\tilde{\mathcal{L}}(d,\hat{d})=
	\begin{cases}
		\mathcal{L}(d,\hat{d}) &\text{if $d=1$}; \\
		\bm{1}_{\hat{d}<2} \cdot \mathcal{L}(2,\hat{d}) &\text{if $d\geq 2$}, \\
	\end{cases}
\end{equation}
where $\bm{1}_{\hat{d}<2}$ is the indicator function that evaluates to $1$ when $\hat{d}<2$. 

For the sake of brevity, $f(\cdot)$ will be used to represent 
the learned embedding function $f(\cdot;\hat{\theta})$ in the subsequent discussion.

\subsection{Deep embedding-based greedy search of codewords}\label{subsec:codeword-search} 
As previously mentioned, single-IDS-correcting codes with large code lengths $n$ are ineffective, 
as the longer the codeword, the higher the likelihood of multiple errors occurring within the same segment. 
Moreover, existing combinatorial codes have already achieved order optimal code rates, 
which are nearly optimal when $n$ is large.

In view of this, the single-IDS-correcting codes with small code lengths $n$ 
and aim to achieve higher code rates are focused. 
By concentrating on smaller code lengths, the random search for a codebook becomes feasible.

A random search-based approach of constructing the code $C(n)$ is repeating the following procedure: 
randomly selecting a sequence from a candidate set 
(initially consists of all possible sequences $A(n)=\{0,1,2,3\}^n$), 
and then filtering out the neighboring sequences of the chosen one from this set.

To outperform the random codeword selecting algorithm in terms of code rate, 
a selecting criterion for choosing codewords from the candidate set 
is crucial in the greedy search procedure. 
Finding a method to select more codewords is equivalent to enhancing the overall code rate. 
In a greedy search approach, codewords with fewer neighbors should be selected in advance. 
However, it remains an open problem to accurately depict the neighbor density of each sequence, 
as only the minimum, maximum, and average sizes of Levenshtein balls with radius $1$ 
have been studied in existing works~\cite{barlev2023size,wang2022size}. 

Fortunately, the deep embedding of Levenshtein distance establishes a connection between
the structural characteristics of Levenshtein distances and the distribution properties of the embedding vectors. 
This deep embedding allows for a rough estimation of the neighboring sequences associated with a given sequence $\bm{s}$
through the Euclidean ball centered at the embedding vector $\bm{u} = f(\bm{s})$. 
By employing a final batch normalization, the embedding model outputs vectors 
that follow a multivariate normal distribution $N(\bm{0},\bm{\Sigma})$ 
with a mean vector $\bm{0}$ and a covariance matrix $\bm{\Sigma}$. 
This probability density function (PDF) of the embedding vectors can then be leveraged 
to evaluate the density of neighbors around the codewords. 
Namely, low-density vectors correspond to sequences that have fewer Levenshtein neighbors. 
By selecting these sequences first, the greedy search makes the efficient choice at each step, 
leaving maximal room for future codewords and thus maximizing the final codebook size. 

Let $m$ denote the dimension of the random vector, 
the PDF of $N(\bm{0},\bm{\Sigma})$ is formulated as 
\begin{equation}\label{eqn:pdf}
	p(\bm{x}) = (2\pi)^{-\frac{m}{2}}|\bm{\Sigma}|^{-\frac{1}{2}}
	\exp\left(-\frac{1}{2}\bm{x}^T\bm{\Sigma}^{-1}\bm{x}\right). 
\end{equation}
The covariance matrix $\bm{\Sigma}$ is easy to estimate with all of the embedding vectors $f(A(n))$. 
Having the estimated covariance matrix $\hat{\bm{\Sigma}}$ in hand, 
the PDF of $N(\bm{0},\hat{\bm{\Sigma}})$ for each embedding vector can be calculated using \eqref{eqn:pdf}. 
To achieve the goal of selecting codewords with fewer neighbors, an effective selecting criterion can be expressed 
in the embedding space as 
the sequence whose embedding vector has the lowest PDF value should be chosen as the codeword in each iteration. 
A step further, by making some simplifications to \eqref{eqn:pdf}, an embedding vector $\bm{u}$ 
and its corresponding sequence should be selected if 
$\bm{u}^T\bm{\Sigma}^{-1}\bm{u}$ is the maximum over the candidate set. 

\begin{algorithm}[htb!]
   \caption{Deep embedding-based greedy search of codewords}
   \label{alg:flowchart}
\begin{algorithmic}[1]
   \Require{Codeword length $n$; the 4-ary alphabet $\Sigma_4$; a pre-trained embedding model $f(\cdot)$.}
   \Ensure{A codebook $C(n)$ where the minimum distance between any two codewords is at least $3$.}
   \State Create the candidate set $A\gets A(n)$ containing all $4^n$ sequences of length $n$.
   \State Initialize an empty codebook $C\gets \emptyset$.
   \State Compute the embedding vectors: $U =\{f(\bm{s}) | \bm{s} \in A(n) \}$. 
   \State Estimate the covariance matrix $\hat{\bm{\Sigma}}$ of the embedding vectors $U$.
   \While{$A \neq \emptyset$}
      \State Select $\bm{s}$ from $A$ that $f(\bm{s})$ have the lowest PDF value: 
      $s=\argmax_{\bm{s}\in A} f(\bm{s})^{T} \hat{\bm{\Sigma}}^{-1} f(\bm{s})$. 
      \State Add $\bm{s}$ to the codebook: $C \gets C \cup \{ \bm{s} \}$. 
      \State Remove neighboring sequences: $A \gets A \backslash B(\bm{s}, 2)$, 
      where $B(\bm{s}, 2) = \{ \bm{s}' \in A | \Delta_{L}(\bm{s}, \bm{s}') \leq 2 \}$. 
   \EndWhile
   \State \Return{$C(n) = C$.}
\end{algorithmic}
\end{algorithm}

The entire procedure for the greedy selecting of codewords is illustrated in \cref{alg:flowchart}. 
Firstly, the statistical distribution $N(\bm{0},\hat{\bm{\Sigma}})$ is estimated on the embedding vectors $f(A(n))$. 
Subsequently, the sequence from the candidate set whose embedding vector possesses 
the lowest PDF value is chosen as a codeword. 
Finally, the candidate set is updated by filtering out the Levenshtein ball, 
and this selecting iteration is repeated until the candidate set becomes empty. 

Once the codebook is generated, users can encode information by choosing codewords 
according to indices from a predefined order. 

\subsection{Deep embedding-based segment correcting}\label{subsec:segment-correction}
Refer a corrupted codeword as a segment. 
In the Levenshtein domain, only brute-force methods are applicable for segment correcting to the best of our knowledge. 
This segment correcting method in BDD involves 
calculating the Levenshtein distance between the segment and all the codewords, 
and then selecting the codeword with the minimal distance as the correction. 
However, this brute-force approach incurs significant computational complexity, 
scaling up to $O(n^2|C(n)|)$. 
It is worth noting that the cardinality of the code $|C(n)|$ grows fast with an increasing $n$ in the experiments. 
While this method can undoubtedly be optimized through techniques like early stopping when calculating Levenshtein distances, 
its complexity remains at least $O(|C(n)|)$.

In this work, the deep embedding vectors and 
their distances are leveraged to correct errors without using the Levenshtein domain. 
The segment correcting procedure is outlined in \cref{fig:segmentCorrect}. 
To be precise, a K-dimension tree (K-d tree)~\cite{bentley1975multidimensional} is constructed from the embedding vectors $f(C(n))$
corresponding to the codewords $C(n)$. 
Subsequently, the embedding vector $\hat{\bm{v}} = f(\hat{\bm{c}})$ of a corrupted segment $\hat{\bm{c}}$ 
is utilized to query its nearest neighbors $\bm{v} = f(\bm{c})$ within this K-d tree, 
thereby confirming the nearest codeword $\bm{c}$ to this segment. 
Significantly, the construction of the K-d tree incurs a one-time complexity cost, 
while the average complexity of querying operations from the K-d tree is $O(\log |C(n)|)$~\cite{friedman1977algorithm}, 
representing a considerable improvement over the previously mentioned brute-force search method. 

\begin{wrapfigure}{r}{0.5\textwidth}
    \vskip -0.3in
	\centering
{\linespread{1}
	\centering
	\tikzstyle{format}=[circle,draw,thin,fill=white]
	\tikzstyle{format_gray}=[circle,draw,thin,fill=gray]
	\tikzstyle{format_rect}=[rectangle,draw,thin,fill=white,align=center,rounded corners=2pt]
	\tikzstyle{arrowstyle} = [->,thick]
	\tikzstyle{network} = [rectangle, minimum width = 3cm, minimum height = 1cm, text centered, draw = black,align=center,rounded corners,fill=green_so,fill opacity=0.5,text opacity=1]
	\tikzstyle{training_batch} = [trapezium, trapezium left angle = 30, trapezium right angle = 150, minimum width = 3cm, text centered, draw = black, fill = cyan_so, fill opacity=0.3,text opacity=1,align=center]		
	\tikzstyle{class_features} = [trapezium, trapezium left angle = 30, trapezium right angle = 150, minimum width = 3cm, text centered, draw = black, fill = cyan_so, fill opacity=0.3,text opacity=1,align=center]
	\tikzstyle{pixel} = [rectangle, draw = black, fill = orange_so, fill opacity=0.5,text opacity=0,align=center]	
	\tikzstyle{pixel_red} = [rectangle, draw = black, fill = red_so, fill opacity=1,text opacity=0,align=center]	
	\tikzstyle{feature} = [rectangle, draw = black, fill = orange_so, fill opacity=0.3,text opacity=0,align=center,rounded corners]	
	\tikzstyle{feature_sfp} = [rectangle, draw = black, fill = violet_so, fill opacity=0.3,text opacity=0,align=center,rounded corners]					
	\tikzstyle{arrow1} = [thick, ->, >= stealth,rounded corners]
	\tikzstyle{arrow1_thick} = [thick, ->, >= stealth, line width=1.2pt,rounded corners]
	\tikzstyle{arrow2} = [thick, dashed, ->, >= stealth]
	\tikzstyle{thick_line} = [thick, line width=1.5pt]
	\tikzstyle{channel} = [fill=white,fill opacity = 0.7, rounded corners=3pt]
	\tikzstyle{channel_shadow} = [fill = gray_so, fill opacity = 0.1, rounded corners]
	\tikzstyle{channel_selected} = [fill = orange_so, fill opacity = 0.5]

	\tikzstyle{dna} = [decoration={coil}, decorate, thick, decoration={aspect=0, segment length=0.5*0.87cm, post length=0.,pre length=0.}]
	\scalebox{0.7} 
	{
	\begin{tikzpicture}[auto,>=latex', thin, start chain=going below, every join/.style={norm}]
		\definecolor{gray_so}{RGB}{88,110,117}
		\definecolor{lightgray_so}{RGB}{207,221,221}
		\definecolor{yellow_so}{RGB}{181,137,0}
		\definecolor{cyan_so}{RGB}{42,161,152}
		\definecolor{orange_so}{RGB}{203,75,22}
		\definecolor{green_so}{RGB}{133,153,0}
		\definecolor{red_so}{RGB}{220,50,47}
		\definecolor{magenta_so}{RGB}{211,54,130}
		\definecolor{violet_so}{RGB}{108,113,196}
		\definecolor{yellow_ad}{RGB}{242,228,201}
		\definecolor{pink_ad}{RGB}{242,182,160}
		\definecolor{green_ad}{RGB}{146,195,185}
		\definecolor{dgreen_ad}{RGB}{104,166,148}
		\definecolor{purple_ad}{RGB}{115,72,88}
		\definecolor{att_blue}{RGB}{185,233,248}
       \definecolor{nature_orange}{RGB}{252,140,98}
       \definecolor{nature_green}{RGB}{102,195,170}
       \definecolor{nature_blue}{RGB}{142,160,204}
       \definecolor{nature_purple}{RGB}{192,187,220}
       \definecolor{nature_lightblue}{RGB}{127,176,210}

       \useasboundingbox  (0,0) rectangle (9.6,6.18);
       
       \scope[transform canvas={scale=1}]
       
       \coordinate (zero) at (0,0);

       \coordinate (upperhalf) at (1,6.18-1.7);
       \coordinate (half) at (1,3.09);
       \coordinate (lowerhalf) at (1,1.7);

       \filldraw[format_rect, fill=nature_green] ($(upperhalf)+(0,0.125*3)$) rectangle ($(upperhalf)+(1.545,0.125*4)$);
       \node at ($(upperhalf)+(0.77,0.21)$) {\small{segment}};
       \filldraw[format_rect, fill=nature_blue, fill opacity=0.5] ($(upperhalf)+(0,-0.125*3)$) rectangle ($(upperhalf)+(1.545,-0.125*2)$);
       \node at ($(upperhalf)+(0.77,-0.55)$) {\small{embed. vec.}};

       \draw[arrow1_thick, dashed] ($(upperhalf)+(0,0.125*3+0.0625)$) -- ($(upperhalf)+(0,0.125*3+0.0625)-(0.4,0)$) -- ($(upperhalf)+(0,-0.125*3+0.0625)-(0.4,0)$) -- ($(upperhalf)+(0,-0.125*3+0.0625)$);
       \node[rotate=90] at ($(upperhalf)+(0,0.0625)+(-0.61,0)$) {\small{$f(\cdot)$}};

       \filldraw[format_rect, fill=nature_orange] ($(lowerhalf)+(0,0.125*3)$) rectangle ($(lowerhalf)+(1.545,0.125*4)$);
       \node at ($(lowerhalf)+(0.77,0.21)$) {\small{correct seq.}};
       \filldraw[format_rect, fill=nature_blue,fill opacity=1] ($(lowerhalf)+(0,-0.125*3)$) rectangle ($(lowerhalf)+(1.545,-0.125*2)$);
       \node at ($(lowerhalf)+(0.77,-0.55)$) {\small{embed. vec.}};
       \draw[format_rect,fill opacity=0,thick,dashed] ($(lowerhalf)+(0,-0.125*3)-(0.125,0.4)$) rectangle ($(lowerhalf)+(1.545,0.125*4)+(0.125,0.125)$);

       \coordinate (upperhalf1) at ($(upperhalf)+(6,0)$);
       \coordinate (half1) at ($(half)+(2,0)$);
       \coordinate (lowerhalf1) at ($(lowerhalf)+(6,0)$);
       \filldraw[format_rect, fill=nature_orange] ($(upperhalf1)+(0,-1.25)$) rectangle ($(upperhalf1)+(1.545,1.25)$);
       \filldraw[format_rect, fill=nature_blue] ($(lowerhalf1)+(0,-1.25)$) rectangle ($(lowerhalf1)+(1.545,1.25)$);
       \node at ($(upperhalf1)+(0.77,1.47)$) {\small{$C(n)$}};
       \node at ($(lowerhalf1)+(0.77,-1.47)$) {\small{$f(C(n))$}};
       \draw[arrow1_thick, dashed] ($(upperhalf1)+(1.545,0)$) -- ($(upperhalf1)+(0.4+1.545,0)$) -- ($(lowerhalf1)+(0.4+1.545,0)$) -- ($(lowerhalf1)+(1.545,0)$);
       \node[rotate=270] at ($0.5*(upperhalf1)+0.5*(lowerhalf1)+(0.61+1.545,0)$) {\small{$f(\cdot)$}};

       \foreach \x/\xtext in {0,...,19}
       {
       \filldraw[format_rect, fill=gray_so, fill opacity=0] ($(upperhalf1)+(0,-1.25+\x*0.25/2)$) rectangle ($(upperhalf1)+(1.545,\x*0.25/2+0.25/2-1.25)$);
       }

       \foreach \x/\xtext in {0,...,19}
       {
       \filldraw[format_rect, fill=gray_so, fill opacity=0] ($(lowerhalf1)+(0,-1.25+\x*0.25/2)$) rectangle ($(lowerhalf1)+(1.545,\x*0.25/2+0.25/2-1.25)$);
       }

       \coordinate (kdtreeDL) at ($(lowerhalf)+(2.445,-0.5)$);
       \coordinate (kdtreeUR) at ($(upperhalf)+(5.1,0.5)$);
       \draw[format_rect,fill opacity=0,thick,dashed] (kdtreeDL) rectangle (kdtreeUR);
       \draw[arrow1_thick] ($(lowerhalf1)$) -- ($(lowerhalf1)+(-0.27,0)$) -- ($(upperhalf1)+(-0.27,0)$) -- ($(kdtreeUR)-(0,0.5)$);
       \node[rotate=90] at ($0.5*(lowerhalf1)+0.5*(upperhalf1)+(-0.46,0)$) {\small{generate K-d tree}};

       \draw[thick,double] ($(upperhalf)+(1.545,-0.125*2.5)$) -- ($(upperhalf)+(0,-0.125*2.5)+(2.445,0)$);
       \draw[arrow1,double] ($(upperhalf)+(0,-0.125*3)+(2.445,0)$) -- ($(upperhalf)+(0,-0.125*3)+(2.445,0)+(-0.27,0)$) -- ($(lowerhalf)+(2.445,0)+(-0.27,-0.1375)$) --($(lowerhalf)+(1.545,0)+(0.125,-0.1375)$);
       \node[rotate=90] at ($0.5*(upperhalf)+0.5*(lowerhalf)+(2.445,0)+(-0.27,-0.1375*0.5)+(0,-0.125*1.5)+(-0.21,0)$) {\small{query neighbors}};

       \coordinate (root) at ($(kdtreeUR)-(1.3275,0.45)$);
       \node at ($(root) + (0,0.7)$) {\small{K-d tree}};
       \coordinate (leftc) at ($(root)+0.5*(-1,-1.732)$);
       \coordinate (rightc) at ($(root)+0.5*(1,-1.732)$);
       \coordinate (lclc) at ($(leftc)+0.5*(-0.6,-1.732)$);
       \coordinate (lcrc) at ($(leftc)+0.5*(0.6,-1.732)$);
       \coordinate (rclc) at ($(rightc)+0.5*(-0.6,-1.732)$);
       \coordinate (rcrc) at ($(rightc)+0.5*(0.6,-1.732)$);

       \coordinate (lclclc) at ($(lclc)+0.5*(-0.3,-1.732)$);
       \coordinate (lclcrc) at ($(lclc)+0.5*(0.3,-1.732)$);
       \coordinate (lcrclc) at ($(lcrc)+0.5*(0.3,-1.732)$);
       \coordinate (lcrcrc) at ($(lcrc)+0.5*(-0.3,-1.732)$);
       \coordinate (rclclc) at ($(rclc)+0.5*(-0.3,-1.732)$);
       \coordinate (rclcrc) at ($(rclc)+0.5*(0.3,-1.732)$);
       \coordinate (rcrclc) at ($(rcrc)+0.5*(0.3,-1.732)$);
       \coordinate (rcrcrc) at ($(rcrc)+0.5*(-0.3,-1.732)$);

       \draw (root) -- (leftc);
       \draw (root) -- (rightc);
       \draw (leftc) -- (lclc);
       \draw (leftc) -- (lcrc);
       \draw (rightc) -- (rclc);
       \draw (rightc) -- (rcrc);
       \draw (lclc) -- (lclclc);
       \draw (lclc) -- (lclcrc);
       \draw (lcrc) -- (lcrclc);
       \draw (lcrc) -- (lcrcrc);
       \draw (rclc) -- (rclclc);
       \draw (rclc) -- (rclcrc);
       \draw (rcrc) -- (rcrclc);
       \draw (rcrc) -- (rcrcrc);

       \draw[format_gray,fill=nature_purple] (root) circle (0.15);
       \draw[format_gray,fill=nature_blue] (leftc) circle (0.15);
       \draw[format_gray,fill=nature_blue] (rightc) circle (0.15);
       \draw[format_gray,fill=nature_blue] (lclc) circle (0.15);
       \draw[format_gray,fill=nature_blue] (lcrc) circle (0.15);
       \draw[format_gray,fill=nature_blue] (rclc) circle (0.15);
       \draw[format_gray,fill=nature_blue] (rcrc) circle (0.15);
       \draw[format_gray,color=white] (lclclc) circle (0.15);
       \draw[format_gray,color=white] (lclcrc) circle (0.15);
       \draw[format_gray,color=white] (lcrclc) circle (0.15);
       \draw[format_gray,color=white] (lcrcrc) circle (0.15);
       \draw[format_gray,color=white] (rclclc) circle (0.15);
       \draw[format_gray,color=white] (rclcrc) circle (0.15);
       \draw[format_gray,color=white] (rcrclc) circle (0.15);
       \draw[format_gray,color=white] (rcrcrc) circle (0.15);

       \node at ($(lclclc)+(0,-0.15)$) {$\vdots$};
       \node at ($(lclcrc)+(0,-0.15)$) {$\vdots$};
       \node at ($(lcrclc)+(0,-0.15)$) {$\vdots$};
       \node at ($(lcrcrc)+(0,-0.15)$) {$\vdots$};
       \node at ($(rclclc)+(0,-0.15)$) {$\vdots$};
       \node at ($(rclcrc)+(0,-0.15)$) {$\vdots$};
       \node at ($(rcrclc)+(0,-0.15)$) {$\vdots$};
       \node at ($(rcrcrc)+(0,-0.15)$) {$\vdots$};

       \endscope
	\end{tikzpicture}
       }	
}
\caption{\label{fig:segmentCorrect} 
Flowchart for the deep embedding-based segment correcting. 
A K-d tree is constructed using the embedding vectors $f(C(n))$ of all the codewords. 
The embedding vector of a segment is used to query the K-d tree for its neighboring sequences. 
}
\vskip -0.6in
\end{wrapfigure}

It is important to recognize that this neighbor-searching procedure is conducted within the embedding space, 
and is based on an approximation rather than the exact Levenshtein distance. 
As a result, the query results from the K-d tree may not always accurately represent the true minimum Levenshtein distances. 
To mitigate this, multiple nearest neighbors can be queried, 
and the Levenshtein distances between the segment and the queried neighboring codewords 
can be double-checked to improve the reliability of the results. 

\section{Experiments and Results}\label{sec:results}
\subsection{Codewords in the embedding space}

\begin{figure*}[htb!]
    \vskip -0.in
    \centering
    \subcaptionbox{deep embedding-based search\label{subfig:a}}[0.32\linewidth]
        {\includegraphics[width=1\linewidth]{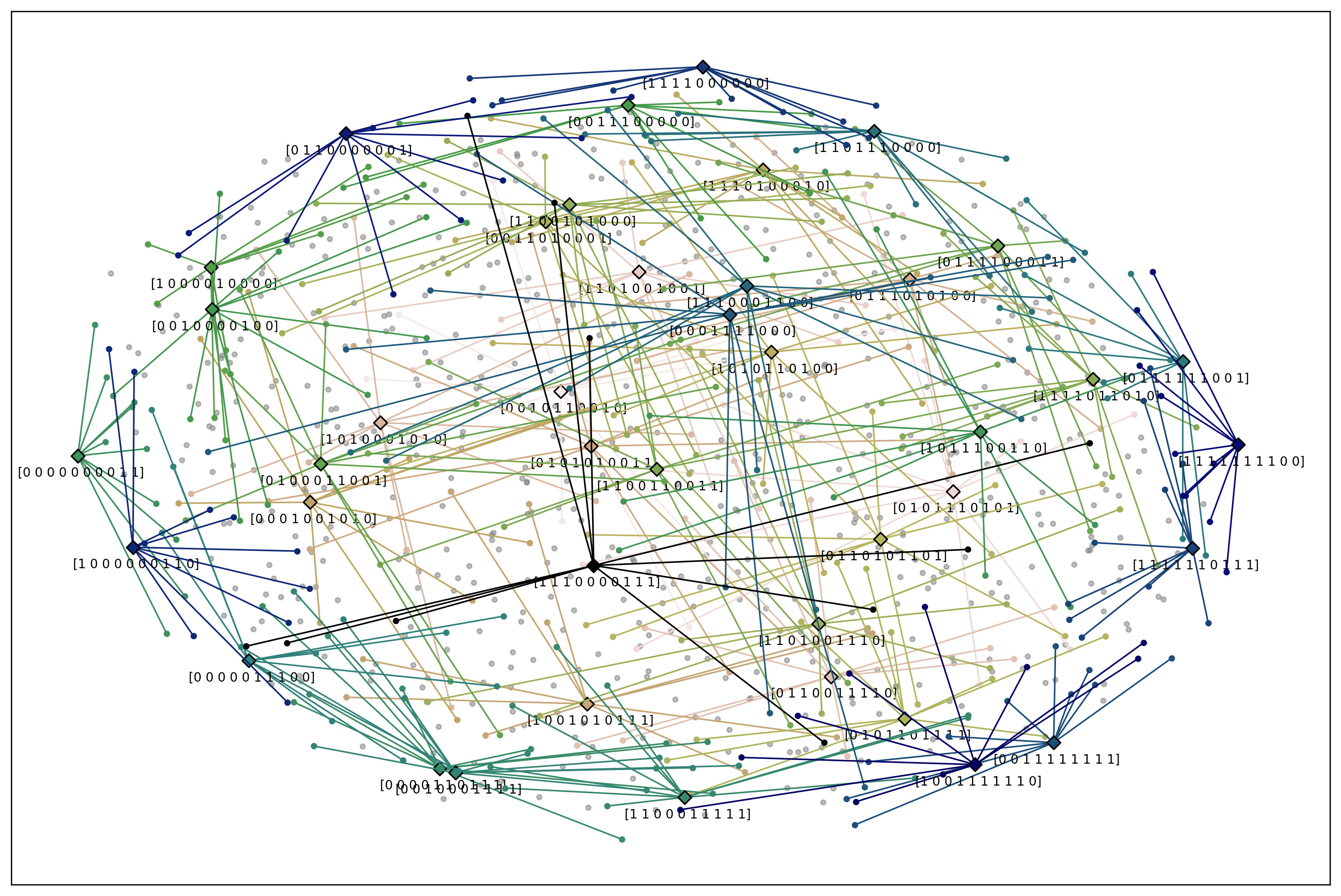}}
    \subcaptionbox{random search\label{subfig:b}}[0.32\linewidth]
        {\includegraphics[width=1\linewidth]{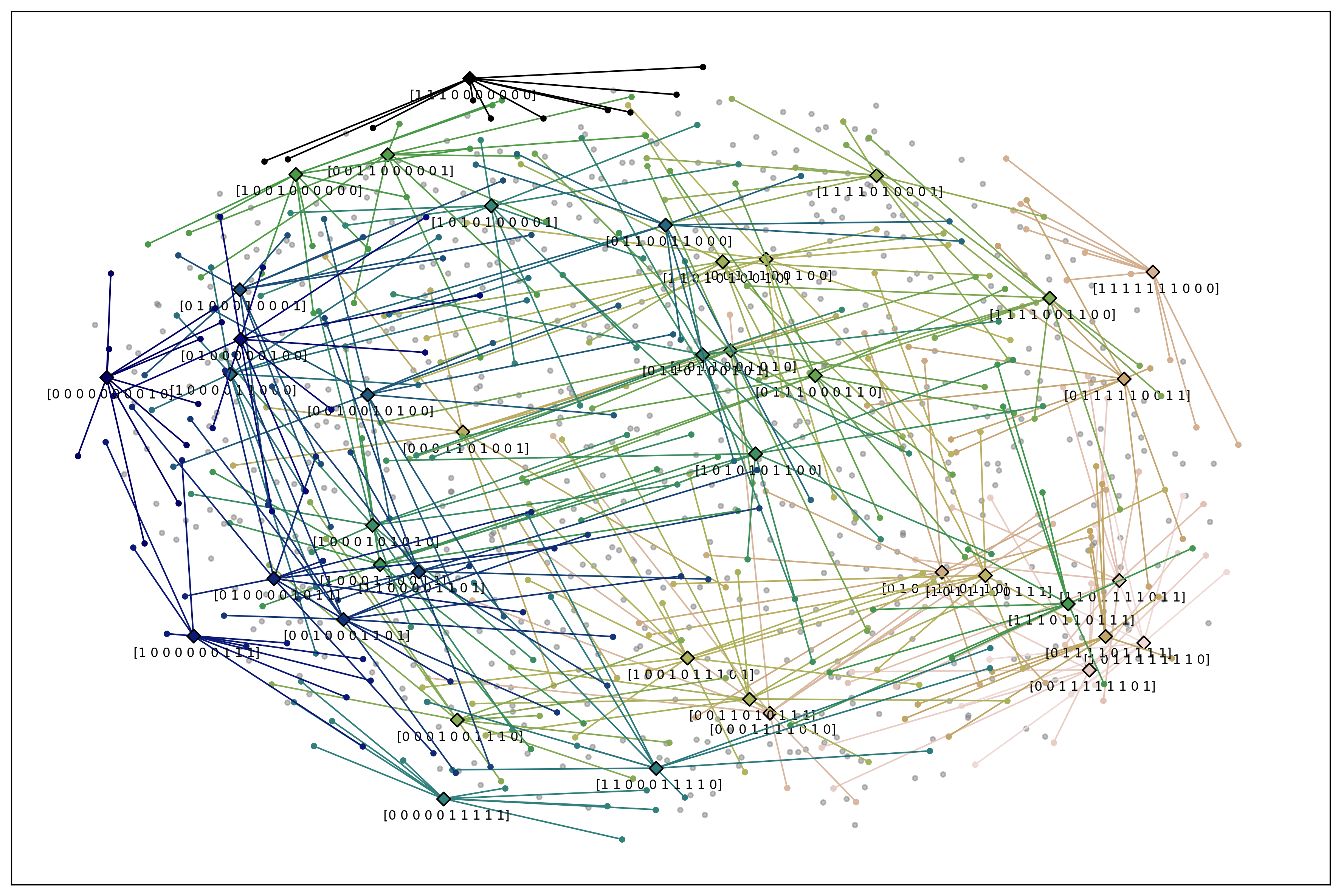}}
    \subcaptionbox{VT code\label{subfig:c}}[0.32\linewidth]
        {\includegraphics[width=1\linewidth]{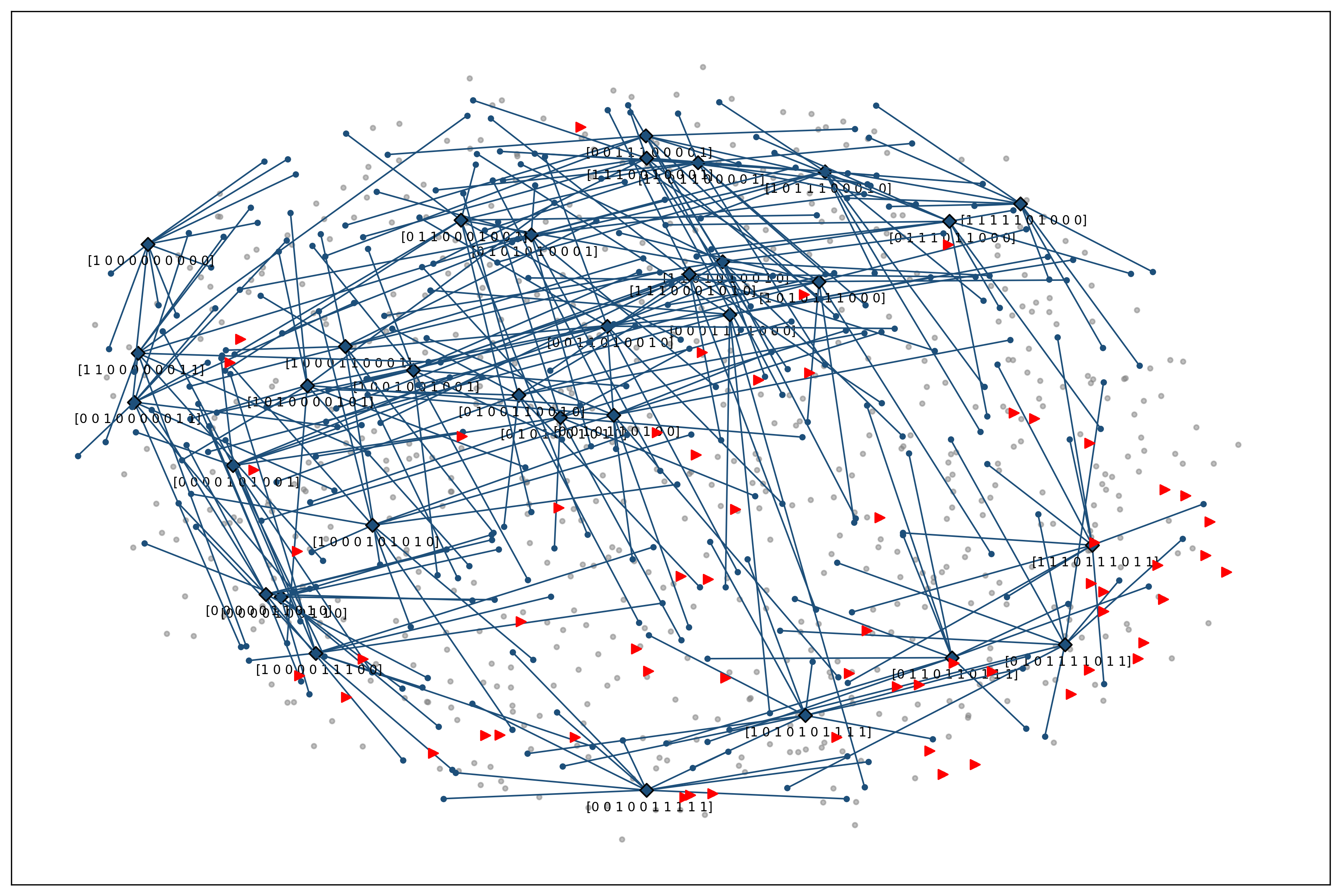}}
    \caption{The relationships between the codewords and their neighboring sequences in the embedding space 
    under three different scenarios:
    (a) results from the proposed deep embedding-based codeword search; 
    (b) results from a random codeword search; 
    (c) results using codewords from the VT code. 
    In each subplot, 
    the diamond markers represent the codewords, 
    and the solid lines connect the codewords to their distance $1$ neighbors. 
    In (a) and (b), 
    the color indicates the order in which the codewords were selected, 
    with darker colors signifying codeword selected earlier. 
    In (c), the red triangle markers identify sequences that are neither codewords nor within a distance of 2 to any codeword. 
    }
    \label{fig:emb}
\vskip -0.in
\end{figure*}

To demonstrate the effectiveness of the proposed deep embedding-based greedy search of 
codewords and the reasonability of using an estimated distribution to 
evaluate the density of the sequences, 
the embedding vectors for all the candidate sequences and the selected codewords 
are visualized and presented in \cref{fig:emb}. 

To enhance the readability, \cref{fig:emb} is generated from experiments with a simplified setting: 
the codeword length is set to $N=10$, the code is 
reduced from a $4$-ary alphabet to a binary alphabet, and the embedding dimension is reduced to $8$. 
To visualize the embedding vectors effectively, the $t$-SNE~\cite{van2008visualizing} is employed 
to project the high-dimensional embedding vectors 
into $\mathbb{R}^2$. The \cref{subfig:a}, \cref{subfig:b}, and \cref{subfig:c} are plotted by the codewords/vectors obtained 
from the proposed deep embedding-based codeword search, a random codeword search, and the VT code, respectively. 
In each subfigure, the diamond markers denote the codewords/vectors, 
and the solid lines connect these to their distance-1 neighbors. 

In \cref{subfig:a} and \cref{subfig:b}, the color scheme indicates the order in which the codewords were selected, 
with darker colors representing codewords selected earlier in the search procedure. 
A comparison of these two subfigures, which represent the projections of the embedding vectors, shows that 
the deep embedding-based search tends to select codewords closer to the periphery of 
the estimated distribution of embedding vectors, 
while the random search selects codewords without any specific pattern. 
This observation aligns with that the embedding vectors follow a multivariate normal distribution, 
and the vectors that deviate from the mean should be selected earlier in \cref{subfig:a} due to their lower PDF 
in the estimated distribution. 

In \cref{subfig:c}, 
the combinatorially constructed codewords from the VT code are plotted over the embedding vectors. 
The red triangle markers identify the isolated sequences that are neither codewords 
nor within a distance of $2$ from any codeword, 
which isolated sequences would be eliminated during the greedy codeword search in the first two subfigures. 
The presence of these isolated sequences suggests 
that the Levenshtein balls with a radius $2$, centered on the VT codewords, 
cannot make a complete coverage on the Levenshtein domain. 
This indicates that a modification of the VT code, which considers these isolated sequences, 
could achieve a larger code rate. 
Additionally, it is also observed in \cref{subfig:c} 
that the VT codewords have a biased distribution in the embedding space used in this experiment, 
with the codewords tending to be located in the North-West, 
while the isolated sequences are more likely found in the South-East of the plane. 

\subsection{Code rate and optimality}
To illustrate the utilization of the deep embedding-based greedy search yields an augmented count of codewords, 
thereby promoting the overall code rate. 
Comparisons are made between the proposed code and the state-of-the-art combinatorial codes. 

Given that the codebook, once generated, is relatively independent of the deep learning model, 
the code with the maximum cardinality from among $10$ runs of the computational experiments is selected. 
The corresponding code rates are calculated using the formula
\begin{equation}
r(n) = \frac{\log_4{|C(n)|}}{n}.
\end{equation}

\subsubsection{DoDo-Code outperforms the combinatorial codes. }
\begin{wrapfigure}{r}{0.5\textwidth}
    \vskip -0.2in
    \centering
    \includegraphics[width=0.9\linewidth]{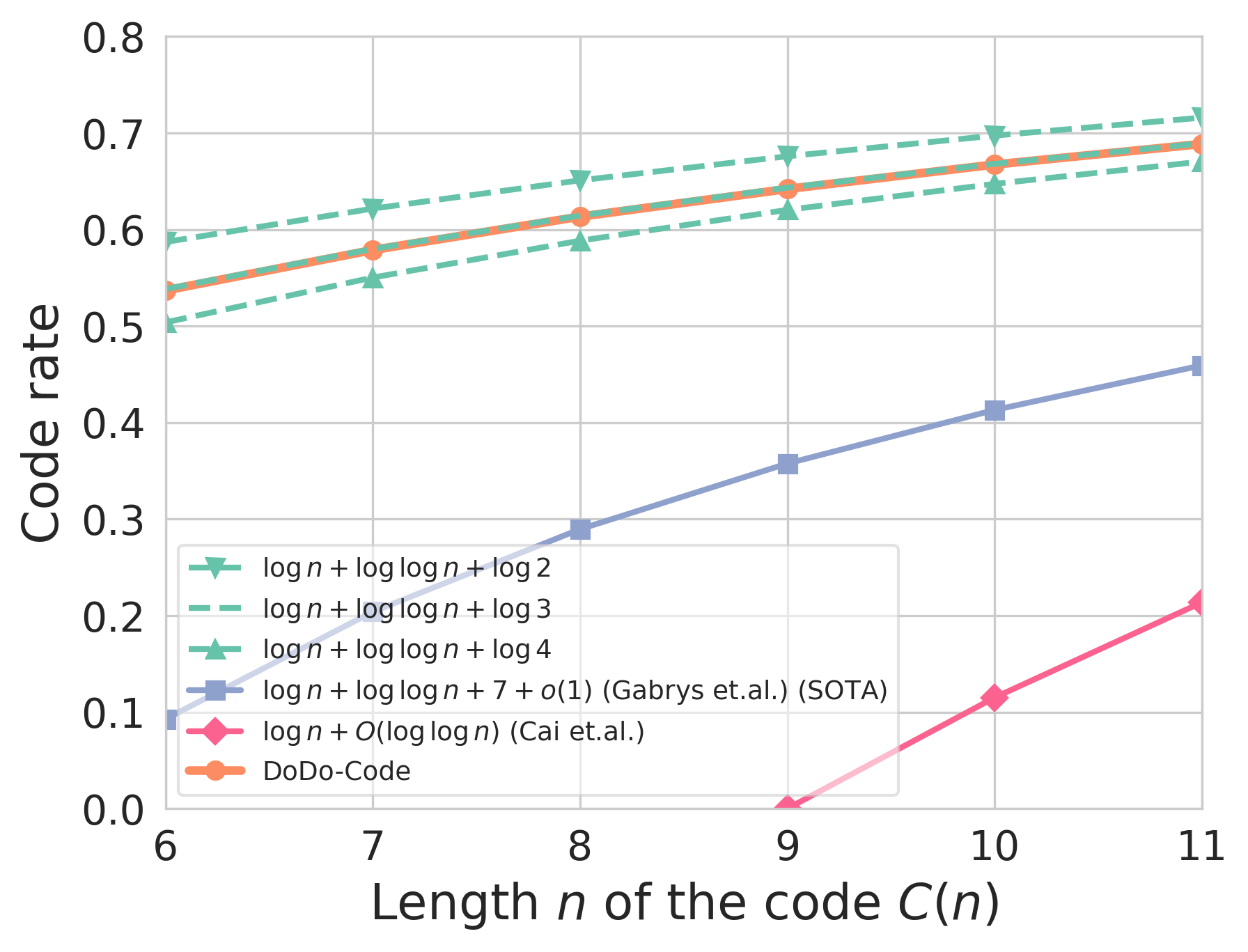}
    \caption{The code rate of code searched by deep embedding-based greedy search strategy, reported as the best in $10$ runs. 
Comparison methods are the state-of-the-art approaches of Cai and Garbys. 
The code rates corresponding to order-optimal redundancies of 
$\log{n} + \log\log{n}+1$, $\log{n} + \log\log{n}+\log 3$, 
and $\log{n} + \log\log{n}+2$ are also plotted. These levels of redundancies are not achieved by any codes before this work. 
}
    \label{fig:coderate}
    \vskip -0.5in
\end{wrapfigure}

The resulting code rates are visualized against the code length in \cref{fig:coderate}. 
For comparison, the code rates of the combinatorial code introduced in~\cite{cai2021correcting} 
which is order-optimal with redundancy of $\log{n} + O(\log\log{n})$ bits, 
as well as the state-of-the-art code rates from~\cite{garbys2023beyond} 
which is also order-optimal using $\log{n} + \log\log{n}+7+o(1)$ redundancy bits, are presented. 
Additionally, the code rates corresponding to the imaginary redundancies of 
$\log{n} + \log\log{n}+1$, $\log{n} + \log\log{n}+\log 3$, 
and $\log{n} + \log\log{n}+2$ in \cref{fig:coderate} are also drawn. 
It's worth noting that for small code lengths $n$, 
no existing $4$-ary code achieves these levels of redundancies. 

Focusing solely on the proposed code, \cref{fig:coderate} clearly shows a trend of increasing code rates 
with longer codeword lengths $n$. 
Notably, for $n=11$, the code rate reaches $68.9\%$. 
Compared to existing state-of-the-art works~\cite{cai2021correcting,garbys2023beyond}, 
the DoDo-Code achieves a significantly higher code rate. 
This improvement is attributed to the fact that, 
although these established combinatorial codes are order optimal, 
the constant terms in their redundancies dominate when $n$ is small, thereby reducing the code rates. 

\subsubsection{DoDo-Code may represent the minimal redundancy achievable. }
Furthermore, when compared to the imaginary redundancies of $\log{n} + \log\log{n}+\{\log2, \log3, \log4\}$
the code rate curve of DoDo-Code lies between $\log{n} + \log\log{n}+1$ and $\log{n} + \log\log{n}+2$ 
and overlaps the curve of $\log{n} + \log\log{n}+\log 3$ exactly. 
It may be claimed that the proposed DoDo-Code is approximately optimal 
using $\log{n} + \log\log{n}+\log 3$ redundancy bits. 
However, this assertion lacks theoretical proof in this study. 

It's worth noting that $\log{n} + \log\log{n}+\log 3$ can be reformulated as $\log{3n} + \log\log{n}$, 
which is nearly the best code rate for a single-IDS-correcting code. 
It is known that correcting a one-bit flip in $N$ bits requires at least $\log{N}$ redundancy bits for information to 
identify the error position. 
In the context of a $4$-ary code, correcting a substitution in $n$ bases requires information on both the error position and 
the substituted letter, which could be any of the other three letters. 
As a result, a minimum of $\log{3n}$ redundancy bits is required. 
Given this, the redundancy $\log{3n} + \log\log{n}$ is very close to its lower bound 
and possibly represents the minimal redundancy achievable. 
Mathematicians might explore this topic further by leveraging combinatorial or probabilistic methods. 

\subsection{Ablation study on embedding space searching and revised PNLL loss}
To illustrate the effectiveness of searching for codewords in the embedding space, 
the comparison is made between the proposed embedding space search and the random codeword 
selecting, which is also introduced for the first time to the best of our knowledge. 
The cardinalities of the codebooks were compared, both in terms of average and maximum results across $10$ runs, 
with the findings presented in \cref{tab:card}. 
The results clearly indicate that using the PDF of the distribution of embedding vectors as the selecting criterion 
in the greedy search yields larger codebooks. 
Furthermore, it is observed that the increase in codeword count becomes more 
pronounced as the codeword length $n$ increases. 
For instance, when $n=11$, the deep embedding-based greedy search identifies $16.8\%$ more 
codewords compared to the random codeword selecting approach. 

For an ablation study on using the revised PNLL loss in \cref{eqn:revisedpnll} 
as the optimization target for the Levenshtein distance embedding network, 
experiments were also conducted engaging the original PNLL loss from \cref{eqn:pnll},
with the results presented in the row labeled DEGS* in \cref{tab:card}. 
The results suggest that employing the revised PNLL loss slightly increased the number of searched codewords.

\begin{table*}[htb!]
    \centering
    \caption{The cardinality of constructed codebook. 
    The results are reported as the mean value and maximum value over $10$ runs of the experiments. 
    The method ``Rand'' stands for random codeword selecting method, the method ``DEGS'' (resp. ``DEGS*'') stands for 
    the proposed deep embedding-based greedy search with the revised PNLL loss (resp. original PNLL loss), and the ``$\Delta$'' stands for the differences. 
    }\label{tab:card}
    {\small
    \begin{tabular}{llrrrrr}
    \toprule
                         & Method       & $n=7$     & $n=8$     & $n=9$     & $n=10$     & $n=11$   \\ \midrule
\multirow{4}{*}{avg.}    & Rand         & 251.5 $\pm$ 5.1     & 813.2 $\pm$ 3.7     & 2694.0 $\pm$ 15.2    & 9091.7 $\pm$ 18.8     & 31071.9 $\pm$ 40.5  \\
                         & DEGS*        & 264.5 $\pm$ 3.3     & 873.3 $\pm$ 8.5     & 2963.5 $\pm$ 18.4    & 10199.4 $\pm$ 57.6    & 35720.4 $\pm$ 297.7  \\
                         & DEGS         & 267.5 $\pm$ 4.7     & 884.8 $\pm$ 15.3    & 3001.6 $\pm$ 8.5     & 10325.4 $\pm$ 48.6    & 35973.1 $\pm$ 157.8  \\
                         & $\Delta$     & $+6.4\%$  & $+8.8\%$  & $+11.4\%$ & $+13.6\%$  & $+15.8\%$ \\\midrule
\multirow{4}{*}{max.}    & Rand         & 259       & 820       & 2717      & 9124       & 31142    \\
                         & DEGS*        & 270       & 887       & 2983      & 10283      & 36191    \\
                         & DEGS         & 275       & 900       & 3011      & 10414      & 36368    \\
                         & $\Delta$     & $+6.2\%$  & $+9.8\%$  & $+10.8\%$ & $+14.1\%$  & $+16.8\%$ \\\bottomrule
    \end{tabular}}
\end{table*}

\subsection{Success rate and experimental time complexity of segment correcting}
The proposed deep embedding-based segment correcting is proposed as an alternative to the 
neighboring search procedure of BDD. 
Experiments were conducted to demonstrate that the proposed method is both reliable and efficient. 

It is worth noting that the searched codewords maintain a minimum mutual Levenshtein distance of $3$, 
ensuring that a single error in a codeword can be confidently corrected by the BDD. 
However, the deep embedding of the Levenshtein distance introduces approximation error, 
which can affect the reliability of the nearest neighbors identified through the tree search in the embedding space. 
A compromise solution is to increase the number $k$ of searched neighbors, 
and then perform a double confirmation using the Levenshtein distances to these $k$ neighbors. 
Experiments with different values of $k$ were conduceted, and the number of failed corrections out of $10^8$ attempts
is presented in \cref{tab:failedcorrection}. 

As indicated in \cref{tab:failedcorrection}, the ratio of failed corrections 
ranges from $0.3\%$ to $0.9\%$ along with different code length, 
when only one embedding vector ($k=1$) is queried. 
When the search is expanded to $k=2$ neighbors, the number of failed corrections decreases significantly. 
Further increasing the searched neighbors to $k=4$, the number of failed corrections is $0$ 
out of correcting $10^8$ modified codewords. 

\begin{table}[htb!]
    \centering
    \caption{Number of failed segment correctings in $10^8$ tries by using tree search. 
    The segments are obtained by randomly one edit modification 
    on the codewords. $k$ is the number of neighbors queried in the tree search. }
    \label{tab:failedcorrection}
    {\small
    \begin{tabular}{lrrrrr}
    \toprule
        & $n=7$ & $n=8$ & $n=9$ & $n=10$ & $n=11$ \\\midrule
        $k=1$ & 328,142   & 398,439   & 465,080   & 740,468    & 828,885    \\
        $k=2$ & 0     & 4,411     & 2,330     & 7,060      & 10,869     \\
        $k=3$ & 0     & 0     & 754     & 0      & 121      \\
        $k=4$ & 0     & 0     & 0     & 0      & 0      \\
        $k=5$ & 0     & 0     & 0     & 0      & 0      \\
    \bottomrule
    \end{tabular}
    }
\end{table}

\begin{wrapfigure}{r}{0.5\textwidth}
    \vskip -0.2in
        \centering
        \includegraphics[width=1\linewidth]{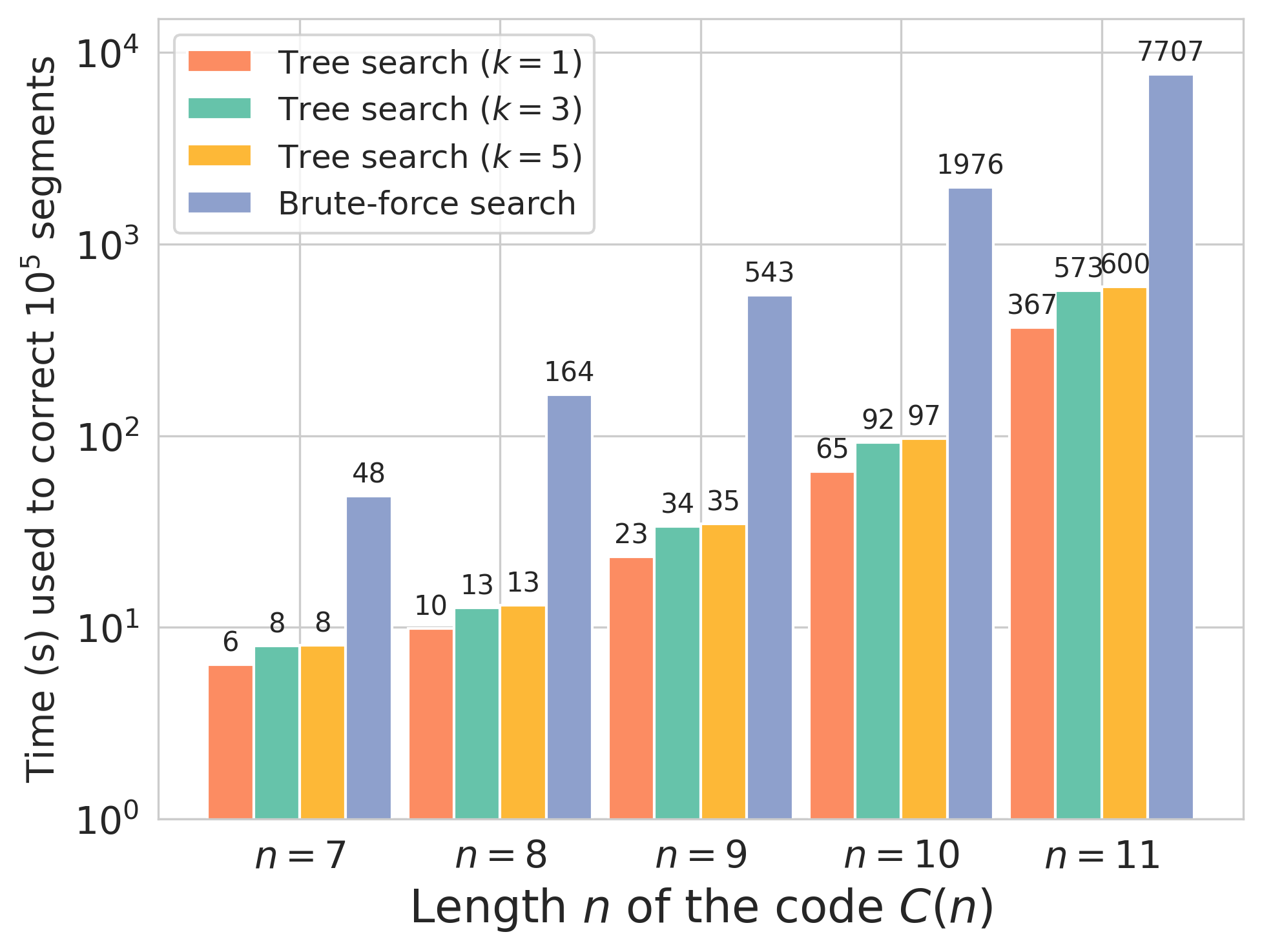}
        \caption{The time used to correct $10^5$ segments. The segments are obtained by randomly one edit modification 
on the codewords. $k$ is the number of neighbors in the tree search. 
The brute-force search calculates the Levenshtein distances 
until the finding of a $1$-distance codeword. The $y$-axis is in $\log$ scale. }
        \label{fig:complexity}
    \vskip -0.3in
\end{wrapfigure}

The proposed deep embedding-based segment correcting utilizes a K-d tree search in Euclidean space, 
which theoretically offers a lower average complexity of $O(\log{|C(n)|})$. 
To demonstrate the efficiency of this method, experiments varying the number $k$ of searched neighbors were performed, 
compared with the brute-force search method, 
which corrects segments by identifying the codeword with the minimal Levenshtein distance. 
As shown in \cref{fig:complexity}, 
the proposed method in Euclidean space significantly reduces time complexity by orders of magnitude
compared to the brute-force approach. 
Moreover, the extra burden of raising $k$ from $1$ to $5$ is minimal, as indicated in \cref{fig:complexity}. 

\subsection{Complexity}\label{subsec:complexity}
When the complexity of copying a codeword is disregarded, 
the encoder of the DoDo-Code operates with negligible complexity, 
since the codebook is pre-generated, and encoding simply consists of selecting a codeword by its index. 

The embedding model, which is implemented by a CNN architecture, maps the sequences to their embedding vectors 
with a complexity of $O(n)$. 
Without considering the one-time cost of building the K-d tree by the embedding vectors of the codebook, 
the segment correcting process incurs a time complexity of at most $O(n)$ 
when querying $k=1$ neighboring sequence. 
The set $A(n)$ containing all sequences of length $n$ over the $4$-ary alphabet 
has a cardinality $4^n$, 
and the code $C(n)$ is a subset of $A(n)$. 
The theoretical expected query time for the K-d tree is $O(\log{|C(n)|})$, which simplifies to $O(n)$, 
considering $|C(n)| < 4^n$. 
When querying $k>1$ neighbors, the double-check on Levenshtein distance increases the time complexity 
to $O(n^2)$. 

The memory complexity to store the K-d tree is $O(m|C(n)|)$, 
where $m$ is the dimension of the embedding vectors and $|C(n)|$ is the cardinality of the codebook. 
Since the tree can be generated on-the-fly from the codebook each time the decoder is initialized, 
and the codebook is deterministically generated using a given embedding model and random seed, 
the only persistent storage needed is for the embedding model itself. 

\subsection{Dataset, source code, and model setting}
All the sequences used for training and testing are generated randomly. 
The groundtruth Levenshtein distance is obtained by a $\mathrm{Python}$ module called $\mathrm{Levenshtein}$. 
Therefore, the experiments run independently of any specific dataset and generate the data on their own. 
The source code is available 
in \url{https://github.com/aalennku/DoDo-Code}.
Unless otherwise specified, the embedding model utilizes an architecture of stacking $10$ $1$D-CNNs, 
with the embedding vector dimension set to $64$. 

\section{Conclusion and Limitations}
To address the code rate issue from 
the segmented error-correcting codes, 
the DoDo-Code, which boasts an ``optimal'' code rate at the short code lengths for $4$-ary IDS correcting code, was proposed. 
By leveraging the deep embedding space as a proxy for the complex Levenshtein domain, 
the mathematically unexplored field is bypassed in the code design. 
The fundamental concept of BDD forms the backbone of the proposed code. 
In the embedding space, an efficient codeword searching algorithm was introduced to maximize the codebook. 
Later, the decoding or correcting algorithm was integrated into the Euclidean embedding space by a K-d tree, 
reducing the computational complexity. 
Experiments illustrated the proposed DoDo-Code 
outperforms the state-of-the-art combinatorial codes in code rate when the code length is small. 

\textbf{Limitations.} The DoDo-Code did not provide explicit mathematical rules for description, 
necessitating a deeper understanding of combinatorial underlying principles. 
The codeword searching relies on a greedy search strategy, which can lead to significant complexity 
when attempting to construct the codebook for large code lengths. 
We hope that future research by mathematicians will uncover the underlying principles governing 
codewords and lead to the invention of mathematically defined codes.

\section*{Acknowledgement}
This work was supported by the National Key Research and Development Program of China under Grant 
2020YFA0712100 and 2025YFC3409900, the National Natural Science Foundation of China, 
and the Emerging Frontiers Cultivation Program of Tianjin University Interdisciplinary Center. 

\bibliographystyle{unsrt}
\bibliography{AlanGuo-bio}

\end{document}